\documentclass[11pt,superscriptaddress,aps,prd,preprint]{revtex4}
\usepackage{amsmath}

\makeatletter
\usepackage[T1]{fontenc}
\usepackage{amsmath}
\usepackage{amssymb}
\usepackage{graphicx}
\usepackage{tabularx}

\usepackage{slashed}

\newcommand{\bea}{\begin{eqnarray}}
\newcommand{\eea}{\end{eqnarray}}

\newcommand{\e}{\epsilon}

\begin{document}

\title{Lorentz violation, Gravitoelectromagnetism and Bhabha Scattering at finite temperature}

\author{A. F. Santos}\email[]{alesandroferreira@fisica.ufmt.br}
\affiliation{Instituto de F\'{\i}sica, Universidade Federal de Mato Grosso,\\
78060-900, Cuiab\'{a}, Mato Grosso, Brazil}

\author{Faqir C. Khanna\footnote{Professor Emeritus - Physics Department, Theoretical Physics Institute, University of Alberta\\
Edmonton, Alberta, Canada}}\email[]{khannaf@uvic.ca}
\affiliation{Department of Physics and Astronomy, University of Victoria,\\
3800 Finnerty Road Victoria, BC, Canada}

\begin{abstract}

Gravitoelectromagnetism (GEM) is an approach for the gravitation field that is described using the formulation and terminology similar to that of electromagnetism. The Lorentz violation is considered in the formulation of GEM that is covariant in its form. In practice such a small violation of the Lorentz symmetry may be expected in a unified theory at very high energy. In this paper a non-minimal coupling term, which exhibits Lorentz violation, is added as a new term in the covariant form. The differential cross section for Bhabha scattering in the GEM framework at finite temperature is calculated that includes Lorentz violation. The Thermo Field Dynamics (TFD) formalism is used to calculate the total differential cross section at finite temperature. The contribution due to Lorentz violation is isolated from the total cross section. It is found to be small in magnitude. 
\end{abstract}


\maketitle

\section{Introduction}

Gravitation is the weakest force in nature, although it is the dominant force in the large scale universe. The theory of Gravity is classical by its origin while other fundamental forces describing microscopic aspects of nature are quantum mechanical. There are several attempts to unify gravity with forces in the Standard Model. The search to unify gravitation and electromagnetism has a long history. The first studies were carried out by Faraday \cite{Faraday} and then by Maxwell \cite{Maxwell}, Heaviside \cite{Heaviside1}, \cite{Heaviside2}, Weyl \cite{Weyl}, Kaluza-Klein \cite{KK}, among others. A formal analogy between the gravitational and the electromagnetic fields led to the notion of Gravitoelectromagnetism (GEM) to describe gravitation. GEM is based on the profound analogy between the Newton's law for gravitation and Coulomb's law for electricity. There are also studies that are based on Einstein's General Relativity (GR) and focus on gravitoelectromagnetism. For example, Lense-Thirring effect showed that in GR a rotating massive body creat a gravitomagnetic field \cite{LT}. The GEM theory emerges from the Einstein theory of gravity (GR) in the linear approximation, i.e., $g_{\mu\nu}=\eta_{\mu\nu}+h_{\mu\nu}$, where $h_{\mu\nu}$ is the perturbation to the linear order. The main structure of GEM emerges as given later in Eq. (1) to Eq. (4). The GEM potential is related to $h_{\mu\nu}$ \cite{Mashhon}. In addition the structure of GEM has a close relationship to the theory of electromagnetism. 

There are three different ways to construct GEM theory: (1) based on the similarity between the linearized Einstein and Maxwell equations \cite{Mashhon}; (2) based on an approach using tidal tensors \cite{Filipe} and (3) based on the decomposition of the Weyl tensor into gravito-magnetic (${\cal B}_{ij}=\frac{1}{2}\epsilon_{ikl}C^{kl}_{0j}$) and gravito-electric (${\cal E}_{ij}=-C_{0i0j}$) components \cite{Maartens}.  A Lagrangian formulation for GEM has been developed \cite{Khanna} using the Weyl tensor approach. GEM allows scattering processes with gravitons as an intermediate state like the photon for electromagnetic scattering. The theory of Gravitoelectromagnetism has been extended from a theory of classical gravity to a quantized theory \cite{Khanna} that allows a perturbative approach to calculating phenomenon in gravity. These do provide a reasonable results for some areas of gravity. In contrast the theory developed by Fierz and Pauli \cite{FP} was for a massive spin-2 field on flat space-time. However, this theory suffered from dangerous pathological, such as the impossibility of the existence of a good massless limit, among others, and was discarded later when a new approach to the gravity field was advanced. This paper is devoted to the study of gravitational Bhabha scattering using the Lorentz-violating framework of GEM.

Lorentz violation can emerge in models unifying gravity with quantum physics such as string theory \cite{Samuel}. Tiny violations of Lorentz and CPT symmetries may be detected experimentally at the Planck scale, $\sim 10^{19}\, \mathrm{GeV}$. The study of Lorentz violation as an extension of the Standard Model (SM) has been undertaken. The Standard Model Extension (SME) is an extensive theoretical framework that includes SM and all possible operators that break Lorentz symmetry \cite{SME1}, \cite{SME2}. The SME is divided into two parts: (i) a minimal extension which has operators with dimensions $d\leq 4$ and preserves conventional quantization, hermitian property, gauge invariance, power counting renormalization, and positivity of the energy and (ii) a non-minimal version of the SME associated with operators of higher dimensions.

Another interesting way to investigate Lorentz violation is to modify the interaction vertex, i.e., a new non-minimal coupling term added to the covariant derivative. The non-minimal coupling term may be CPT-odd or CPT-even. There are some applications with this new interaction term, such as: its effect on the cross section of the electron-positron scattering has been investigated \cite{Casana1}, modification in the Dirac equation in the non-relativistic regime has been analyzed \cite{Casana2},  radiative generation of the CPT-even gauge terms of the SME have been constructed \cite{Casana3},  the CPT-even aether-like Lorentz-breaking term has been generated in the extended Lorentz-breaking QED \cite{Petrov1}, \cite{Petrov2}, effects induced on the magnetic and electric dipole moments have been investigated \cite{Casana4}, Lorentz violation in Bhabha scattering in Electromagnetic theory at finite temperature has been studied \cite{Our2}, among others. In this paper the Bhabha scattering in the non-minimal coupling framework for GEM is analyzed at finite temperature. The Lorentz-violating parameter belongs to the gravity sector of SME and has dimension five, i.e. it is a part of the non-minimal version of SME. The temperature in stars would indicate the magnitude of total contribution of the Lorentz violating term to the cross section. It is important to note that there is a similar term in the non-minimal version of the electromagnetism sector of the SME. All this shows similarity between these two theories. The Thermo Field Dynamics (TFD) formalism is used to introduce finite temperature effects in order to estimate variation of the cross section for GEM.

TFD is a thermal quantum field theory  \cite{Umezawa1}, \cite{Umezawa2}, \cite{Umezawa22}, \cite{Khanna1}, \cite{Khanna2} formalism. Its basic elements are: (i) the doubling of the original Fock space, composed of the original and a fictitious space (tilde space) and (ii) the Bogoliubov transformation that is a rotation of these two spaces. The original and tilde space are related by a mapping, tilde conjugation rules. The physical variables are described by non-tilde operators. As a consequence, the propagator is written in two parts: $T = 0$ and $T\neq 0$ components. TFD is a natural formalism to describe systems in equilibrium at finite temperature.
 
This paper is organized as follows. In section II, the GEM theory in its Lagrangian formalism is presented. In section III, the GEM Lagrangian with Lorentz-violating term is considered. In section IV, a brief introduction to TFD formalism is presented. In section V, the differential cross section for Bhabha scattering for GEM including Lorentz-violating parameter at finite temperature is calculated. In section VI, some concluding remarks are presented.

\section{GEM and its lagrangian formalism}

The Gravitoelectromagnetic (GEM) theory describes the dynamics of the gravitational field . In a flat space-time the Maxwell-like equations of GEM are given as
\bea
&&\partial^i{\cal E}^{ij}=-4\pi G\rho^j,\label{01}\\
&&\partial^i{\cal B}^{ij}=0,\label{02}\\
&&\epsilon^{( i|kl}\partial^k{\cal B}^{l|j)}+\frac{\partial{\cal E}^{ij}}{\partial t}=-4\pi G J^{ij},\label{03}\\
&&\epsilon^{( i|kl}\partial^k{\cal E}^{l|j)}+\frac{\partial{\cal B}^{ij}}{\partial t}=0,\label{04}
\eea
where ${\cal E}_{ij}$ is the gravitoelectric field and ${\cal B}_{ij}$ is the gravitomagnetic field that are defined in terms of the Weyl tensor components  ($C_{ijkl}$), i.e., ${\cal E}_{ij}=-C_{0i0j}$ and ${\cal B}_{ij}=\frac{1}{2}\epsilon_{ikl}C^{kl}_{0j}$. Here $G$ is the gravitational constant, $\epsilon^{ikl}$ is the Levi-Civita symbol, $\rho^j$ is the vector mass density and $J^{ij}$ is the mass current density. The symbol $(i|\cdots|j)$ denotes symmetrization of the first and last indices, i.e., $i$ and $j$. 

The GEM fields ${\cal E}$ and ${\cal B}$, with components ${\cal E}^{ij}$ and ${\cal B}^{ij}$, are defined as (details are given in \cite{Khanna})
\bea
{\cal E}&=&-\mathrm{grad}\,\varphi-\frac{\partial \tilde{\cal A}}{\partial t},\\
{\cal B}&=&\mathrm{curl}\,\tilde{\cal A},
\eea
where $\tilde{\cal A}$, with components ${\cal A}^{\mu\nu}$, is a symmetric rank-2 tensor field of the gravitoelectromagnetic tensor potential, and $\varphi$ is the GEM vector counterpart of the electromagnetic scalar potential $\phi$. The tensor fields, ${\cal E}_{ij}$ and ${\cal B}_{ij}$, are elements of a rank-3 tensor, the gravitoelectromagnetic tensor, $F^{\mu\nu\alpha}$, defined as
\bea
F^{\mu\nu\alpha}=\partial^\mu{\cal A}^{\nu\alpha}-\partial^\nu{\cal A}^{\mu\alpha},
\eea
where $\mu, \nu,\alpha=0, 1, 2, 3$. The non-zero components of ${F}^{\mu\nu\alpha}$ are ${F}^{0ij}={\cal E}^{ij}$ and ${F}^{ijk}=\epsilon^{ijl}{\cal B}^{lk}$ where $i, j=1, 2, 3$. Using the gravitoelectromagnetic tensor the Maxwell-like equations, eqs. (\ref{01})-(\ref{04}), are written in a covariant form as
\bea
\partial_\mu{F}^{\mu\nu\alpha}&=&4\pi G{\cal J}^{\nu\alpha},\\
\partial_\mu{\cal G}^{\mu\langle\nu\alpha\rangle}&=&0,
\eea
where ${\cal G}^{\mu\nu\alpha}$ is the dual GEM tensor, that is defined as
\bea
{\cal G}^{\mu\nu\alpha}=\frac{1}{2}\epsilon^{\mu\nu\gamma\sigma}\eta^{\alpha\beta}{F}_{\gamma\sigma\beta},
\eea
and ${\cal J}^{\nu\alpha}$ is a rank-2 tensor that depends on the mass density, $\rho^i$, and the current density $J^{ij}$. With these definitions the GEM Lagrangian is written as
\bea
{\cal L}_G=-\frac{1}{16\pi}{F}_{\mu\nu\alpha}{F}^{\mu\nu\alpha}-G\,{\cal J}^{\nu\alpha}{\cal A}_{\nu\alpha}.\label{L_G}
\eea
This Lagrangian formalism is constructed using the symmetric gravitoelectromagnetic tensor potential $A_{\mu\nu}$  as the fundamental field that describes the gravitational interaction. Although $A_{\mu\nu}$  has similar symmetry properties to those of  $h_{\mu\nu}$, which is a tensor defined in Einstein Gravity in the weak field approximation, our approach is different, since the nature of $A_{\mu\nu}$ is quite different from $h_{\mu\nu}$. An essential difference, the tensor potential is connected directly with the description of the gravitational field in flat space-time and it has nothing to do with the perturbation of the space-time metric.

\section{GEM with Lorentz-violating term}

The main objective of this paper is to calculate the differential cross section for Bhabha scattering using the graviton-fermion interaction described by the Lagrangian
\bea
{\cal L}&=&-\frac{1}{16\pi}F_{\mu\nu\alpha}F^{\mu\nu\alpha}-\bar{\psi}\left(i\gamma^\mu \overleftrightarrow{D_\mu}-m\right)\psi,\label{eq1}
\eea
where the first term is the GEM lagrangian and the second term is the Dirac lagrangian. Here  $\psi$ is the fermion field with $\bar{\psi}=\psi^\dagger \gamma_0$, $m$ is the fermions mass, $\gamma^\mu$ are Dirac matrices and $D_\mu$ is the covariant derivative.  To study Lorentz violation effects in the graviton-fermion interaction, the usual covariant derivative is modified by a non-minimal coupling term, i.e.,
\bea
\overleftrightarrow{D_\mu}=\overleftrightarrow{\partial_\mu}-\frac{1}{2}igA_{\mu\nu}\overleftrightarrow{\partial^\nu}+\frac{1}{4}\bigl(k^{(5)}\bigl)_{\mu\nu\alpha\lambda\rho}\gamma^\nu F^{\alpha\lambda\rho},\label{der}
\eea
where $g=\sqrt{8\pi G}$ is the gravitational coupling constant and $\bigl(k^{(5)}\bigl)_{\mu\nu\alpha\lambda\rho}$ is a tensor that belongs to the gravity sector of the non-minimal SME with mass dimension $d=5$ \cite{QG-K}. Then unit of this parameter is given as ${{\rm GeV}^{4-d}}$. Since the action is dimensionless, the lagrangian ${\cal L}$ in eq. (\ref{eq1}) has dimension ${\rm GeV}^4$, then the tensor potential $A_{\mu\nu}$ has dimension ${\rm GeV}^1$. Here the Weyl formulation is used to investigate the flat space-time treatment of the theory of gravitation, i.e. GEM. This formulation is similar to the case of fermions with electromagnetic interactions. The correspondence between Lorentz violation effects for the electromagnetic (EM) field and for the weak field gravitational field, i.e. GEM in the mininal version of SME has been studied \cite{QG}. In this paper the similarity between GEM with Lorentz violation and the non-minimal part of the Electromagnetic sector of SME is utilized. 

Using eq. (\ref{der}) the interaction part of the lagrangian becomes
\bea
{\cal L}_I&=&-\frac{g}{4}A_{\mu\nu}\left(\bar{\psi}\gamma^\mu\partial^\nu\psi-\partial^\mu\bar{\psi}\gamma^\nu\psi\right)-\frac{1}{4}\bigl(k^{(5)}\bigl)_{\mu\nu\alpha\lambda\rho} F^{\alpha\lambda\rho}\bar{\psi}\sigma^{\mu\nu}\psi,
\eea
where $\sigma^{\mu\nu}=\frac{i}{2}\left(\gamma^\mu\gamma^\nu-\gamma^\nu\gamma^\mu\right)$ and by definition $A\overleftrightarrow{\partial^\mu}B\equiv\frac{1}{2}\left(A\partial^\mu B-\partial^\mu AB\right)$.  The first term describes the usual interaction between gravitons and fermions and the second term is a new interaction that leads to Lorentz violation. This new interaction describes the non-minimal coupling between the GEM field and the fermion bilinear. It is similar to the non-minimal coupling between the electromagnetic field and the bilinear fermion field \cite{Kost_H}. Then  the vertices are 
\bea
\bullet&\rightarrow& -\frac{ig}{4}\left(\gamma^\lambda p_1^\rho+p_2^\lambda\gamma^\rho\right)\equiv V^{\lambda\rho}_{(0)}\\
\circ&\rightarrow& -\frac{1}{2}\bigl(k^{(5)}\bigl)^{\mu\nu\alpha\lambda\rho}\sigma_{\mu\nu} q_\alpha\equiv V^{\lambda\rho}_{(1)}.
\eea
Here the momentum transfer, $q_\alpha$ is considered as $q_\alpha=(\sqrt{s},0)$, with $s$ being the center of mass energy.

The main interest of this paper is to study the graviton-fermion interaction at finite temperature. In the next section thermal quantum field theory is introduced.

\section{TFD formalism}

In this section a brief introduction to TFD formalism is considered. TFD is a real time formalism of quantum field theory at finite temperature. This is obtained when a thermal vacuum or ground state, $|0(\beta)\rangle$,  is defined. The thermal average of an observable is given by the vacuum expectation value in an extended Hilbert space. There are two necessary basic ingredients  to construct the TFD formalism: (a) the doubling the degrees of freedom in a Hilbert space and (b) the Bogoliubov transformations. This doubling is defined by the tilde ($^\thicksim$) conjugation rules, associating each operator in $S$ to two operators in $S_T$ , where the expanded space is $S_T=S\otimes \tilde{S} $, with $S$ being the standard Hilbert space and $\tilde{S}$ the fictitious Hilbert space. For an arbitrary operator ${\cal A}$ the standard doublet notation is
\bea
{\cal A}^a=\left( \begin{array}{cc} {\cal A}\\
\xi\tilde{{\cal A}}^\dagger \end{array} \right),
\eea
where $\xi =+1(-1)$ for bosons (fermions). The Bogoliubov transformation introduces a rotation in the tilde and non-tilde variables and thermal quantities. The Bogoliubov transformations are different for fermions and bosons.

Considering fermions with $c_p^\dagger$ and $c_p$ being creation and annihilation operators respectively, in the standard Hilbert space and  $\tilde{c}_p^\dagger$ and $\tilde{c}_p$ being operators in the tilde space.  For fermions the Bogoliubov transformations are
\bea
c_p&=&\mathsf{u}(\beta) c_p(\beta) +\mathsf{v}(\beta) \tilde{c}_p^{\dagger }(\beta), \label{f1}\\
c_p^\dagger&=&\mathsf{u}(\beta)c_p^\dagger(\beta)+\mathsf{v}(\beta) \tilde{c}_p(\beta),\label{f2}\\
\tilde{c}_p&=&\mathsf{u}(\beta) \tilde{c}_p(\beta) -\mathsf{v}(\beta) c_p^{\dagger}(\beta),\label{f3} \\
\tilde{c}_p^\dagger&=&\mathsf{u}(\beta)\tilde{c}_p^\dagger(\beta)-\mathsf{v}(\beta)c_p(\beta),\label{f4}
\eea
where $\mathsf{u}(\beta) =\cos \theta(\beta)$ and $\mathsf{v}(\beta) =\sin \theta(\beta)$. The anti-commutation relations for creation and annihilation operators  are similar to those at zero temperature and are given as
\bea
\left\{c(k, \beta), c^\dagger(p, \beta)\right\}&=&\delta^3(k-p),\nonumber\\
 \left\{\tilde{c}(k, \beta), \tilde{c}^\dagger(p, \beta)\right\}&=&\delta^3(k-p),\label{ComF}
\eea
and other anti-commutation relations are null.

Now consider bosons with $a_p^\dagger$ and $a_p$ being creation and annihilation operators respectively, in the standard Hilbert space and $\tilde{a}_p^\dagger$ and $\tilde{a}_p$ being operators in the tilde space, then the Bogoliubov transformations are
\bea
a_p&=&\mathsf{u}'(\beta) a_p(\beta) +\mathsf{v}'(\beta) \tilde{a}_p^{\dagger }(\beta), \\
a_p^\dagger&=&\mathsf{u}'(\beta)a_p^\dagger(\beta)+\mathsf{v}'(\beta) \tilde{a}_p(\beta),\\
\tilde{a}_p&=&\mathsf{u}'(\beta) \tilde{a}_p(\beta) +\mathsf{v}'(\beta) a_p^{\dagger}(\beta), \\
\tilde{a}_p^\dagger&=&\mathsf{u}'(\beta)\tilde{a}_p^\dagger(\beta)+\mathsf{v}'(\beta)a_p(\beta),
\eea
where $\mathsf{u}'(\beta) =\cosh \theta(\beta)$ and $\mathsf{v}'(\beta) =\sinh \theta(\beta)$. Algebraic rules for thermal operators are
\bea
\left[a(k, \beta), a^\dagger(p, \beta)\right]&=&\delta^3(k-p),\nonumber\\
 \left[\tilde{a}(k, \beta), \tilde{a}^\dagger(p, \beta)\right]&=&\delta^3(k-p),\label{ComB}
\eea
and other commutation relations are null.

An important note, the propagator in TFD formalism is written in two parts: one describes the flat space-time contribution and the other displays the thermal effect. Here our interest is in the graviton propagator at finite temperature, which is given as
\bea
\langle 0(\beta)|\mathbb{T}\left[A_{\mu\nu}(x)A_{\rho\lambda}(y)\right]|0(\beta)\rangle=i\int \frac{d^4k}{(2\pi)^4}e^{-ik(x-y)}\Delta_{\mu\nu\rho\lambda}(k,\beta),\label{prop}
\eea
where $\mathbb{T}$ is the time ordering operator and $\Delta_{\mu\nu\rho\lambda}(k,\beta)=\Delta_{\mu\nu\rho\lambda}^{(0)}(k)+\Delta_{\mu\nu\rho\lambda}^{(\beta)}(k)$ with 
{\small
\bea
\Delta_{\mu\nu\rho\lambda}^{(0)}(k)&=& \frac{\eta_{\mu\rho}\eta_{\nu\lambda}+\eta_{\mu\lambda}\eta_{\nu\rho}-\eta_{\mu\nu}\eta_{\rho\lambda}}{2k^2}\,\tau_0\nonumber\\
\Delta_{\mu\nu\rho\lambda}^{(\beta)}(k)&=&-\frac{\pi i\delta(k^2)}{e^{\beta k_0}-1}\left( \begin{array}{cc}1&e^{\beta k_0/2}\\e^{\beta k_0/2}&1\end{array} \right)(\eta_{\mu\rho}\eta_{\nu\lambda}+\eta_{\mu\lambda}\eta_{\nu\rho}-\eta_{\mu\nu}\eta_{\rho\lambda}),
\eea
}
where $\Delta_{\mu\nu\rho\lambda}^{(0)}(k)$ and $\Delta_{\mu\nu\rho\lambda}^{(\beta)}(k)$ are zero and finite temperature parts respectively and
\bea
\tau_0=\left( \begin{array}{cc}1 & 0 \\ 0 & -1\end{array} \right).
\eea

\section{The differential cross section - Bhabha scattering }

Our interest is to calculate the cross section at finite temperature for the process, $e^-(p_1)+e^+(p_2)\rightarrow e^-(q_1)+e^+(q_2)$, with one graviton exchange including Lorentz violating terms. The Feynman diagrams, that describe this process, are given in FIG. 1. Contribution of Lorentz violation terms make small contribution to the cross section due to GEM theory. Until the Lorentz violation becomes significant, higher order contributions are not expected to be large.
\begin{figure}[h]
\includegraphics[scale=0.3]{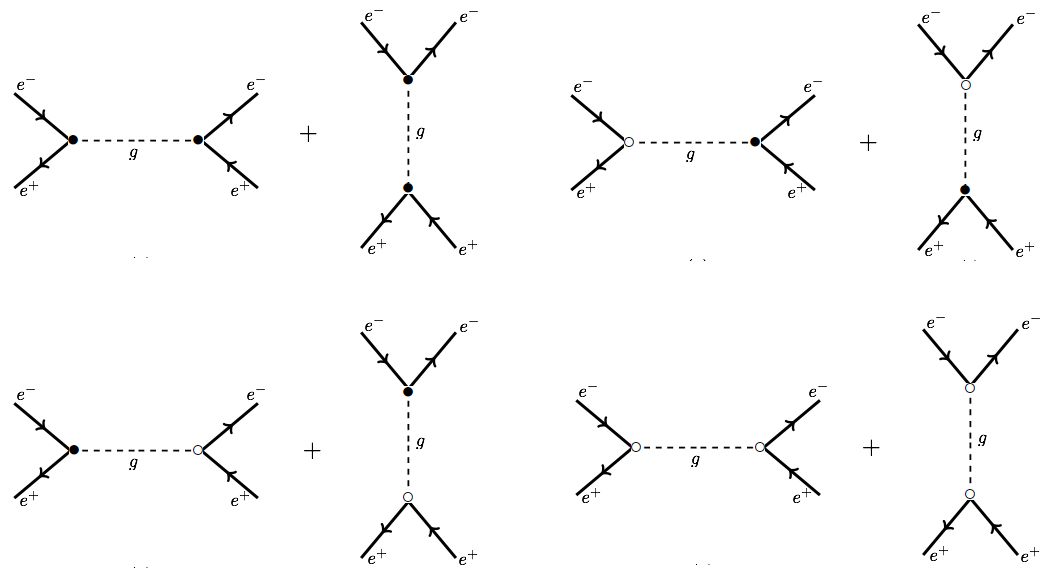}
\caption{GEM Bhabha Scattering with one graviton exchange. Here, $\bullet$ represent the usual GEM vertex and $\circ$ represent the new vertex due to the Lorentz violation.}
\end{figure}

The calculation is carried out in the center of mass frame (CM) where we have
\bea
p_1&=&(E,\vec{p}),\quad\quad p_2=(E,-\vec{p}),\nonumber\\
q_1&=&(E,\vec{p'})\quad\quad\mathrm{and}\quad\quad q_2=(E,-\vec{p'}),
\eea
where $|\vec{p}|^2=|\vec{p'}|^2=E^2$, $\vec{p}\cdot\vec{p'}=E^2\cos\theta$ and
\bea
p_1\cdot p_2=q_1\cdot q_2&=&2E^2, \quad p_1\cdot q_1=E ^2(1-cos\theta)\nonumber\\
p_1\cdot q_2=q_1\cdot p_2&=&2E^2, \quad p_2\cdot q_2=E ^2(1-cos\theta).
\eea

The differential cross section is defined as
\bea
\frac{d\sigma}{d\Omega}=\left(\frac{\hbar^2 c^2}{64\pi^2 s}\right)\cdot\frac{1}{4}\sum_{spins}\bigl|{\cal M(\beta)}\bigl|^2,\label{cs}
\eea
where $s$ is the CM energy, ${\cal M}(\beta)$ is the S-matrix element at finite temperature.  In addition an average over the spin of the incoming particles and summing over the spin of outgoing particles is included.

The transition amplitude for GEM Bhabha scattering is calculated as
\bea
{\cal M}(\beta)=\langle f,\beta| \hat{S}^{(2)}| i,\beta\rangle,
\eea
with $\hat{S}^{(2)}$, the second order term, of the $\hat{S}$-matrix that is defined as
\bea
\hat{S}&=&\sum_{n=0}^\infty\frac{(-i)^n}{n!}\int dx_1dx_2\cdots dx_n \mathbb{T} \left[ \hat{{\cal L}}_{I}(x_1) \hat{{\cal L}}_{I}(x_2)\cdots \hat{{\cal L}}_{I}(x_n) \right],
\eea
where $\hat{{\cal L}}_{I}(x)={{\cal L}}_{I}(x)-\tilde{{\cal L}}_{I}(x)$ describes the interaction. The thermal states are
\bea
| i,\beta\rangle&=&c_{p_1}^\dagger(\beta)d_{p_2}^\dagger(\beta)|0(\beta)\rangle, \nonumber\\
 | f,\beta\rangle&=&c_{p_3}^\dagger(\beta)d_{p_4}^\dagger(\beta)|0(\beta)\rangle ,
\eea
with $c_{p_j}^\dagger(\beta)$ and $d_{p_j}^\dagger(\beta)$ being creation operators. The transition amplitude becomes
\bea
{\cal M}(\beta)&=&\frac{(-i)^2}{2!}\int d^4x\,d^4y\langle f,\beta|({\cal L}_I{\cal L}_I-\tilde{\cal L}_I\tilde{\cal L}_I)| i,\beta\rangle\nonumber\\
&=&\Bigl({\cal M}_0(\beta)+{\cal M}_\kappa(\beta)+{\cal M}_{\kappa\kappa}(\beta)\Bigl)-\left(\tilde{\cal M}_0(\beta)+\tilde{\cal M}_\kappa(\beta)+\tilde{\cal M}_{\kappa\kappa}(\beta)\right),
\eea
where ${\cal M}_0(\beta)$ is the matrix element of the Lorentz invariant, ${\cal M}_\kappa(\beta)$ is the linear term in the Lorentz violation and ${\cal M}_{\kappa\kappa}(\beta)$ is the second order in the Lorentz-violating parameter. This last term will be ignored since its contribution is of the fourth order in the Lorentz-violating parameter, then is very small when compared with the contribution of the ${\cal M}_\kappa(\beta)$ term. An important note, there are similar equations for matrix elements that include tilde operators.

The fermion field is written as
\bea
\psi(x)=\int dp\, N_p\left[c_p u(p)e^{-ipx}+d_p^\dagger v(p)e^{ipx}\right],
\eea
where $c_p$ and $d_p$ are annihilation operators for electrons and positrons, respectively, $N_p$ is the normalization constant, and $u(p)$ and $v(p)$ are Dirac spinors. The Lorentz invariant part of the transition amplitude becomes
{\small
\bea
&&{\cal M}_0(\beta)=-\frac{ig^2}{16} N\int d^4x\,d^4y\,\int d^4p(\mathsf{u}^2-\mathsf{v}^2)^2\langle 0(\beta)|\mathbb{T}[A_{\mu\nu}(x)A_{\rho\lambda}(y)]|0(\beta)\rangle\nonumber\\
&\times& \Bigl[\bar{u}(q_1)(\gamma^\mu p_1^\nu+q_1^\mu\gamma^\nu) u(p_1)\bar{v}(p_2)(\gamma^\rho p_2^\lambda +q_2^\rho\gamma^\lambda) v(q_2)e^{-ix(p_2-p_1)}e^{iy(q_2-q_1)}\nonumber\\
&-&\bar{u}(q_1)(\gamma^\mu q_1^\nu+p_1^\mu\gamma^\nu) v(p_1)\bar{v}(q_2)(\gamma^\rho q_2^\lambda +p_2^\rho\gamma^\lambda) u(p_2)e^{ix(q_1+p_1)}e^{-iy(q_2+p_2)} \Bigl],
\eea}
where Bogoliubov transformations eqs. (\ref{f1})-(\ref{f4}) are used. With $\mathsf{u}(\beta) =\cos \theta(\beta)$ and $\mathsf{v}(\beta) =\sin \theta(\beta)$ we get $(\mathsf{u}^2-\mathsf{v}^2)^2= \tanh^2(\frac{\beta |k_0|}{2})$, where $k_0=\omega$. Using the graviton propagator definition at finite temperature, given in eq. (\ref{prop}), and the definition of the four-dimensional delta function,
\bea
&&\int d^4x\,d^4y\,e^{-ix(p_2-p_1+k)}e^{-iy(q_1-q_2-k)}=\delta^4(p_2-p_1+k)\delta^4(q_1-q_2-k),
\eea
the transition amplitude is written as 
\bea
{\cal M}_0(\beta)&=&-\frac{ig^2}{16}\Bigl[\bar{u}(q_1)(\gamma^\mu p_1^\nu+q_1^\mu\gamma^\nu) u(p_1)D_{\mu\nu\rho\lambda}(p_1-q_1)\bar{v}(p_2)(\gamma^\rho p_2^\lambda +q_2^\rho\gamma^\lambda) v(q_2)\nonumber\\
&-& \bar{u}(q_1)(\gamma^\mu q_1^\nu+p_1^\mu\gamma^\nu) v(p_1)D_{\mu\nu\rho\lambda}(q_1+p_1)\bar{v}(q_2)(\gamma^\rho q_2^\lambda +p_2^\rho\gamma^\lambda) u(p_2))\Bigl]\nonumber\\
&\times&\tanh^2\Bigl(\frac{\beta E_{CM}}{2}\Bigl),
\eea
where $|(p_1-q_1)_0|=|(q_1+p_1)_0|=E_{CM}$ has been used, with $E_{CM}$ being the energy of the CM and
\bea
D_{\mu\nu\rho\lambda}(k)\equiv \Delta(k)\,(\eta_{\mu\rho}\eta_{\nu\lambda}+\eta_{\mu\lambda}\eta_{\nu\rho}-\eta_{\mu\nu}\eta_{\rho\lambda})
\eea
with 
\bea
\Delta(k)&=&\frac{1}{k^2}\left( \begin{array}{cc}1 & 0 \\ 
0 & -1\end{array} \right)-\frac{2\pi i\delta(k^2)}{e^{\beta k_0}-1}\left( \begin{array}{cc}1&e^{\beta k_0/2}\\e^{\beta k_0/2}&1\end{array} \right).\label{delta}
\eea

In a similar way the linear term in the Lorentz violating parameter becomes
\bea
{\cal M}_\kappa(\beta)&=&\frac{g}{4}\Bigl[\bar{u}(q_1)(\gamma^\mu p_1^\nu+q_1^\mu\gamma^\nu) u(p_1)D_{\mu\nu\rho\lambda}(p_1-q_1)\bar{v}(p_2) V^{\lambda\rho}_{(1)} v(q_2)\nonumber\\
&-& \bar{v}(q_1)(\gamma^\mu q_1^\nu+p_1^\mu\gamma^\nu) u(p_1)D_{\mu\nu\rho\lambda}(q_1+p_1)\bar{u}(p_2) V^{\lambda\rho}_{(1)} v(p_2))\Bigl]\nonumber\\
&\times&\tanh^2\Bigl(\frac{\beta E_{CM}}{2}\Bigl).
\eea

For evaluating the differential cross section the relevant quantity is $|{\cal M}|^2=\sum{\cal M}{\cal M}^*$, where the sum is over spins. Then
\bea
\bigl|{\cal M}(\beta)\bigl|^2=\bigl|{\cal M}_0(\beta)+{\cal M}_\kappa(\beta)\bigl|^2.
\eea
This calculation is accomplished using the completeness relations:
\bea
\sum_{spins} u(p_1)\bar{u}(p_1)&=&\slashed{p}_1+m, \nonumber\\
\sum_{spins} v(p_1)\bar{v}(p_1)&=&\slashed{p}_1-m.
\eea
In addition, the relation
\bea
\bar{v}(p_2)\gamma_\alpha u(p_1)\bar{u}(p_1)\gamma^\alpha v(p_2)=\mathrm{tr}\left[\gamma_\alpha u(p_1)\bar{u}(p_1)\gamma^\alpha  v(p_2)\bar{v}(p_2)\right]
\eea
is used. Henceforth the electron mass is ignored since all the momenta are much larger than the electron mass, i.e., ultra relativistic limit. 

Then the differential cross section at finite temperature becomes
\bea
\left(\frac{d\sigma}{d\Omega}\right)_T&=&\frac{g^4E^8}{4096\pi^2 s}\Bigl\{\Delta_1^2\left(1952\cos\theta+460\cos 2\theta+32\cos 3\theta+\cos 4\theta+1651\right)\nonumber\\
&+&32\Delta_2^2\left(\cos 2\theta+3\right)+128\Delta_1\Delta_2\left(\cos\theta+7\right)\cos^4(\theta/2)\nonumber\\
&+&\frac{32 (k^5)^2}{g^2}\bigl[-\Delta_1^2\Bigl(\cos^2\theta(1-\cos\theta)+24(1-\cos\theta)^3+10(1-\cos\theta)^2(\cos\theta-1)\Bigl)\nonumber\\
&-&120\Delta_2^2+\Delta_1\Delta_2\Bigl(2(10+9\cos\theta)+20(1+\cos\theta)\Bigl)\bigl]\Bigl\}\tanh^4\left(\frac{\beta E_{CM}}{2}\right),
\eea
where $\Delta_1\equiv\Delta_1(p_1-q_1)$ and $\Delta_2\equiv\Delta_2(p_1+q_1)$ are defined from eq. (\ref{delta}) and are written explicitly as
\bea
\Delta_1&=&\frac{1}{(p_1-q_1)^2}\left( \begin{array}{cc}1 & 0 \\ 
0 & -1\end{array} \right)-\frac{2\pi i\delta((p_1-q_1)^2)}{e^{\beta (p_1-q_1)_0}-1}\left( \begin{array}{cc}1&e^{\beta (p_1-q_1)_0/2}\\e^{\beta (p_1-q_1)_0/2}&1\end{array} \right)
\eea
\bea
\Delta_2&=&\frac{1}{(p_1+q_1)^2}\left( \begin{array}{cc}1 & 0 \\ 
0 & -1\end{array} \right)-\frac{2\pi i\delta((p_1+q_1)^2)}{e^{\beta (p_1+q_1)_0}-1}\left( \begin{array}{cc}1&e^{\beta (p_1+q_1)_0/2}\\e^{\beta (p_1+q_1)_0/2}&1\end{array} \right).
\eea
Here it is considered that the beam is perpendicular to the background, i.e., $\bigl(k^{(5)}\bigl)_{\mu\nu\alpha\lambda\rho}\,p^\rho=0$.

At zero temperature limit, $\tanh^4\left(\frac{\beta E_{CM}}{2}\right)\rightarrow  1$, $\Delta_{1}\rightarrow  \frac{i}{2(p_1-q_1)^2}$ and $\Delta_{2}\rightarrow  \frac{i}{2(p_1+q_1)^2}$. Then the differential cross section is
\bea
\left(\frac{d\sigma}{d\Omega}\right)&=&\left(\frac{d\sigma}{d\Omega}\right)_{GEM}\Biggl[1+\frac{\bigl(k^{(5)}\bigl)^2}{g^2{\cal O}}\Bigl(128\cos^2\theta\sin^6(\theta/2) -4924\cos\theta + 1552 \cos 2\theta\nonumber\\ 
&-& 630 \cos 3\theta +140\cos 4\theta +14\cos 5\theta+ 3876\Bigl)\Biggl],\label{odd1}
\eea
where ${\cal O}\equiv 1864\cos\theta+540\cos 2\theta+56\cos 3\theta+5\cos 4\theta+1631$. Here  $\left(\frac{d\sigma}{d\Omega}\right)_{GEM}$ is the differential cross section for the GEM field \cite{AFS}, Lorentz invariant case, and is given by
\bea
\left(\frac{d\sigma}{d\Omega}\right)_{GEM}=-\frac{g^4E^4}{n\,\pi^2\,s}\frac{\left(1864\cos\theta+540\cos 2\theta+56\cos 3\theta+5\cos 4\theta+1631\right)}{(\cos\theta-1)^2},
\eea
with $n$ being a numerical factor defined as $n\approx6,6\times 10^4$.

It is clear that results at finite temperature are likely to be small, but these may be measurable in some cases. This will provide the role of Lorentz breaking components of the transition operators.

\section{Conclusions}

The standard model of particle physics and Electromagnetic field have been Lorentz covariant at all energies so far. In addition it is believed that the same will be the case for Gravitational field like Einstein theory and Gravitoelectromagnetic field. The gravitational field is considered to have its presence to a much larger time scale. A particular question to ask is: has Lorentz invariance been valid for all times for systems in a gravitational field or in a field consistent with the quantum fields? This has led to consideration of validity of this invariance. Then the question is posed to consider consequences of violation of Lorentz covariance. The present study is directed to investigate the role of temperature in such a violation. The Lorentz violation at finite temperature is studied for gravitoelectromagnetism (GEM). GEM is a gravitational theory obtained from the Einstein field equations in the linear approximation. And as stated earlier GEM has close resemblance to the electromagnetic theory. The formalism of Thermofield dynamics is used to calculate the differential cross section of gravitons at finite temperatures in the presence of Lorentz violation. It is well known  that GEM field with Lorentz violation is similar to the electromagnetic field in the non-minimal version of SME. The present study gives a brief look at this aspect and the possible expectations in results in experiments. It is conceivable that interior of stars may show some results that will corroborate or discount the presence of Lorentz violation in the gravitational field and possibly in the case of standard model. Our results show that the differential scattering cross section for Bhabha scattering depends on temperature and details are presented. Variation of scattering cross section with Lorentz violation term in the starting Lagrangian presents the question about its impact on the cross section when the temperature is changing, such as in the interior of stars. This would have impact of different magnitude depending on temperature. This will help us to understand the role of Lorentz violation term depending on the nature of star with its internal temperature.
In addition, our results are calculated in the CM frame. However the coefficients in the CM frame are not constant because all experiments with beams involve non-inertial laboratories on the Earth, which is rotating in the standard Sun-centered inertial frame (SCF). Then CM coefficients need to be converted to SCF coefficients as discussed in \cite{Kost_H}, \cite{Kost2002}, \cite{Kost1998}. 

\section*{Acknowledgments}
It is a pleasure to thank V. A. Kosteleck\'y for useful remarks about the Lorentz-violating coefficient for GEM field.

\end{document}